\documentclass[12pt]{article}
\usepackage{graphics}
\usepackage{epsfig}
\usepackage{graphicx}
\usepackage{graphicx}
\begin{document}

\begin{center}
{\Large\bf Bound States in the Continuum in Elasticity} 

\vskip 0.3cm
{\bf O. Haq}${}^{1*}$ and {\bf S.~V.~Shabanov}${}^2$

\vskip 0.3cm
${}^1$ {\it Department of Physics, University 
of Florida, Gainesville, FL 32611, USA}\\
${}^2$ {\it Department of Mathematics, University 
of Florida, Gainesville, FL 32611, USA}\\
${}^*$Corresponding author:\ \ omerhaq1@ufl.edu

\end{center}

\begin{abstract}
Diffraction of elastic waves 
is considered for a system consisting of two 
parallel arrays of thin (subwavelength) cylinders that are arranged periodically.
The embedding media supports waves with all polarizations, one longitudinal
and two transverse, having different dispersion relations. 
An interaction with scatters mixes longitudinal and one of the transverse modes.  
It is shown that the system supports bound states 
in the continuum (BSC) that have no specific polarization, that is, there 
are standing waves localized in the scattering structure whose wave numbers lies
in the first open diffraction channels for both longitudinal and transverse 
modes. 
BSCs are shown to exists only for specific distances between 
the arrays and for specific values of the wave vector component along the array.  
An analytic solution is obtained
for such BSCs. For distances 
between the parallel arrays much larger than the wavelength,  
BSCs is proved to exist due to destructive interference of the far field resonance radiation, similar to 
the interference in a Fabry-Perot interferometer, that can occur
simultaneously for both propagating modes.   
 
\end{abstract}
 \subsubsection{\label{sec:level1}Introduction}

 Bound States in the Continuum (BSC) have been studied in the context of many physical system from photonics to quantum mechanics, since the seminal paper by von Neumann and Wigner \cite{NW1,NW2}.  Examples of BSC have been explored in just about every field of wave physics: quantum mechanics, electromagnetism, and acoustics.
Systems supporting BSC can be viewed as resonators with infinitely high 
 quality factors. Due to these unusual physical properties BSC 
are intensively studied both theoretically and experimentally, especially 
in photonics  \cite{Hsu} and recently even in plasma-photonic systems \cite{plasmaphotonic}.  In particular,
acoustic BSC have been observed in ``Wake Shedding' experiment'' 
\cite{wakeShedding} as well as in acoustic wave guides with 
obstructions \cite{awg}. A relation between elastic BSC and photonic resonances in opto-mechanical crystal slabs \cite{Optomech} 
as well as surface acoustic waves on anisotropic crystals and non-periodic layered structures \cite{anistropic}-\cite{layered}
has been investigated 
using a group-theoretical approach.  Elastic systems offer a rich playground, both theoretically and experimentally, to investigate BSC especially in view of
mechanical metamaterials \cite{MMM1}--\cite{MMM4} 
with unusual stiffness, rigidity and compressibility.

There is no universal mechanism for formation of BSC, however there are a few classifications which may not be mutually exclusive \cite{Hsu}. 
In some instances, the existence of BSCs can be explained by a destructive 
interference of diffracted waves in the asymptotic region \cite{PRL2008}.
If  a system  of subwavelength scatterers that are arranged 
in a plane happens to be a resonator for incident 
plane waves, then two such systems separated by a distance 
form a Fabry-Perot interferometer with resonating interfaces.
For a large enough distance, the interfaces are interacting only    
through diffracted (propagating) modes, while an interaction via
evanescent modes is suppressed due to an exponential decay of the latter.
Given quality factors of each resonating interface,
it is then not difficult to compute the quality factor 
of the combined structure using the standard Fabry-Perot summation 
of transmitted and reflected waves. 
It appears that the quality factor 
depends on the distance between the arrays and the wave speed in the media
between the interfaces. There exist distances at which 
the quality factor becomes infinite, thus indicating the existence of BSC.
At these distances, the diffracted wave from each interface undergo 
destructive interference in the asymptotic region. In other words,
each of the resonating Fabry-Perot interfaces acts as a mirror for 
a standing wave  that becomes confined between the interfaces despite that
its wave numbers lies in open diffraction channels  \cite{JMP2010}.
 
Many of the photonic applications such as filters and sensor can be translated into the context of elastics with little efforts due to similarities 
of the two theories. However there are drastic differences among the two such as the presence of longitudinally polarized wave modes in elastics as well as the boundary conditions at the interface. Even for simple scattering geometries,
like the aforementioned periodic arrays of cylinders, the normal traction 
boundary condition \cite{Landau} at 
the interface of the elastic rods (which is necessary for a
mechanical equilibrium of the system) leads to a coupling between the longitudinal
(compression) and one of the transverse (sheer) modes. Given that these modes
have different dispersion relations, the Fabri-Perot argument \cite{PRL2008}
becomes inapplicable to prove the existence of elastic BSCs, and a full 
analysis of the scattering problem is required.

In this paper, BSC for elastic waves are investigated in 
a system of two periodic arrays of cylindrical scatters separated by a distance.
Due to the translational
symmetry, transverse waves polarized parallel to the cylinders are decoupled from 
the other two polarization modes (one transverse and one longitudinal).
The latter modes are coupled and have different 
dispersion relations. The problem is solved by means of the Lippmann-Schwinger
formalism in the dipole approximation. 
A confinement of an elastic wave 
between two arrays that couple the transverse and longitudinal 
modes, due to normal traction boundary conditions, requires that both modes interfere destructively in 
the asymptotic region.
It is not surprising that by adjusting the distance one can confine 
a particular polarization mode if it is not coupled to any other mode
(e.g., the longitudinal one). It is remarkable that by adjusting  
geometrical and spectral parameters it is possible to create a standing wave consisting of both coupled polarization modes. The conditions under which
such elastic BSC exists comprise our main result. 
To our knowledge, this is the first analytic 
solution for a BSC containing coupled waves with different dispersion relations.
It is noteworthy that so far only 
electromagnetic BSCs with no specific polarization 
were found in numerical studies \cite{Hsu2}-\cite{Bulgakov} where both transverse modes propagates with the {\it same} group velocity.
Although in the present study the existence of such BSCs is established 
under several simplified assumptions such as a dipole approximation
in solving the scattering problem, 
a large distance between the arrays (to justify the use of the Fabri-Perot
argument), and some simplifying conditions on the elastic properties 
of the scatterers, the stated formalism holds even if these 
assumptions are dropped. However, the analysis of the Lippmann-Schwinger
integral equation becomes mathematically involved and will be presented 
elsewhere.      

\subsubsection{\small Elastic wave scattering on a periodic array of cylinders}    
 
A displacement 
field $u$ in an elastic media has three components, one longitudinal (compression wave) and
two transverse (sheer waves). In general,
the longitudinal and transverse 
components have different group velocities $c_l$ and $c_t$, respectively, determined by the media mass
density $\rho_b$ and lame coefficients, $\lambda_b$ and $\mu_b$.
A scattering structure 
is described by the relative mass and lame coefficients in units of the corresponding
background quantities, denoted here 
by $\xi_{\rho,\lambda,\mu}$. For example, $\xi_\rho =(\rho_s-\rho_b)/\rho_b$,
where $\rho_s$ is the mass density of a scattering structure,
and similarly for the relative lame coefficients,
so that $\xi_{\rho,\lambda,\mu}(r)=0$
at any point $r$ where no scatterers are present.

Let the scattering structure be a periodic array of 
cylinders, that is, $\xi_{\rho,\lambda,\mu}(r)\neq 0$ on cylinders whose 
axes are arranged periodically in a plane and all cylinders   
have the same radius, which is much smaller than the period.
Owing to the translational symmetry along axes of the cylinders,
the field $u$ depends only on two variables spanning the plane 
perpendicular to the cylinders. 
In what follows, a unit system is chosen so that the period of the array is one,
the $z$ axis is chosen to be parallel to the cylinders, and 
the array is periodic along the $y$ axis so that $u$ depends on $x$ and $y$.
The mode $u_z$ is decoupled 
from the two modes in the $xy$ plane due to the translational symmetry.
The scattering problem for $u_z$ is fully analogous to the electromagnetic 
case studied in detail in \cite{JMP2010}. So the existence of elastic BSC for this mode  
can readily be established. 
The compression mode and the sheer mode polarized in the $xy$ plane
remains coupled through the scattering structure and, yet, they have
different dispersion relations. The objective is 
to investigate whether elastic BSCs consisting of the two coupled modes exist in
a system of two parallel arrays of periodically positioned cylinders.  

Let $u_j(r)$, $j=x,y$, be the amplitude of a monochromatic elastic (displacement) field of frequency $\omega$ at a point $r=(x,y)$;
the indices $i$, $j$, and $k$ are used to denote the $x$ and $y$ components of a vector in the coordinate system 
described above, while the indices $a$ and $b$ 
to indicate polarization states in the $xy$ plane,
$a=l$ (longitudinal) and $a=t$ (transverse).
For example,
the incident wave is $u^0_{a,j}(r)=u^{0}_{a} e^{ik_{a,x}x+ik_yy}\hat{e}_{a,j}$ where 
the unit polarization vector $\hat{e}_{a}$ is parallel to 
 the wave vector $(k_{a,x},k_y)$ for the longitudinal wave and perpendicular to it
for the transverse one, the length of $(k_{a,x},k_y)$ for each 
polarization mode is set by the dispersion relation,
$c_{a}^{2}k^2_{a}=c_{a}^{2}(k_{a,x}^{2}+k_{y}^{2})= \omega^{2}$, thus defining $k_{a,x}$ via $\omega$, $u^{0}_{a}$ corresponds to the incident amplitude of polarization $a$. The standard basis $\hat{x}$, $\hat{y}$ can always be 
converted to the basis $\hat{e}_a$ by a suitable rotation. 

The elastic field is a superposition of the incident and scattered fields, $u_j(r)=u^{0}_j(r)+u^{S}_j(r)$.  
Adopting the Einstein rule 
for summation over repeated indices,  
the governing equations for $u_j(r)$ have the form \cite{Landau}:
\begin{eqnarray}
\label{eq:1}
&&[(\omega^{2}+c_{t}^{2}\triangle)\delta_{jk}+(c_{l}^{2}-c_{t}^{2})\nabla_j 
\nabla_k] u_k(r)=\frac{1}{\rho_{b}}P_j(r)\\
\label{eq:2}
&&P_j(r)=-\rho_{b}(\omega^{2}\xi_{\rho}(r)u_j(r)+\nabla_k \sigma_{kj}(r))
\\
\nonumber
&&\sigma_{kj}(r)=(c_l^2-2c_t^2)\xi_{\lambda}(r)\nabla_{i} u_{i}(r) \delta_{kj}
+c_t^2\xi_{\mu}(r) (\nabla_k u_{j}(r)+\nabla_j u_{k}(r))\,.
\end{eqnarray}
The incident wave $u^0_{j}(r)$ satisfies the homogeneous equation (\ref{eq:1})
 with $P_{j}(r)=0$. Without loss of generality, put $\rho_{b}=1$.
The quantity $P_j(r)$ has a simple physical interpretation
as an induced dipole moment per unit area in the $xy$ plane. 
The scattered field at a position $r$, $du^{S}_{j}(r)$, is a result
the radiation field produced by an infinitesimal induced dipole moment centered at position $r_{0}$, $P_{i}(r_{0})d^2r_0$; it has the standard form
 $$
 du^{S}_j(r)=G_{ji}(r-r_{0})P_i(r_{0})d^{2}r_{0}
 $$
where $G_{ji}(r)$ is the 
Green's function for the differential operator 
in the left side of (\ref{eq:1}) satisfying Sommerfeld radiation conditions
at spatial infinity:
$$
G_{ji}(r)= \frac{i}{4c_{l}^{2}}\left(\,
\frac{\nabla_j \nabla_i}{k_{l}^{2}}\,\right) H_{0}^{(1)}(k_{l} \mid r \mid)-\frac{i}{4c_{t}^{2}}\left(\frac{\nabla_j \nabla_i}{k_{t}^{2}}+
\delta_{ji}\right)H_{0}^{(1)}(k_{t} 
\mid r \mid)
$$
where $k_{a}=\frac{\omega}{c_{a}} $ 
are the wave numbers of the longitudinal and transverse modes, and 
$H_0^{(1)}$ is the Hankel function of the first kind. 
It follows that the total displacement field reads
\begin{equation}\label{eq:LS}
 u_j(r)=u^{0}_j(r)+\int_{\Omega} G_{ji}(r-r_{0})P_i(r_{0})d^{2}r_{0} \,,
\end{equation}
 where $\Omega$ is the region occupied by scatterers.

To simplify the discussion further
and avoid complicated technicalities associated with solving the 
Lippmann-Schwinger integral
equation (\ref{eq:LS}),
the simplest case with two open diffraction channels, one transverse and one longitudinal, is considered so that the frequency range is limited to
\begin{eqnarray}\label{eq:opendc}
c_{l}^{2}k_{y}^{2}<&\omega^{2}&<c_{t}^{2}(k_{y}-2\pi)^{2}\equiv \omega_1^2\,,\\ \nonumber
0<&k_{y}&<\frac{2\pi \alpha}{1+\alpha}\,,
\end{eqnarray}
where $k_y$ is the $y$ component of the wave vector, and
the materials are assumed isotropic, in this case,
$\alpha=c_{t}/c_{l}<1/\sqrt{2}$ \cite{Landau}. 
Every linearly independent wave state in this system is uniquely described by 
a pair of spectral parameters $(\omega^2,k_y)$ and its polarization state $a$.
The theory is symmetric 
under $k_{y}\rightarrow-k_{y}$ so that $k_{y}>0$ without 
loss of generality. The stated upper bound on $k_{y}$ is necessary in order
for the longitudinal continuum edge to lie below the first  diffraction threshold for the transverse mode. The special case $k_{y}=0$ will be discussed later. The relative density, $\xi_{\rho}=\frac{\rho_{s}-\rho_{b}}{\rho_{b}}=\rho_{s}-1$, is required to be positive. 
Since the density in elastics plays the same role as the permittivity in electromagnetism, the positivity of $\xi_\rho$ is required for the existence 
of BSC. A negative relative density corresponds to a repulsive potential in quantum mechanics or conductors in electromagnetism (in either case,
BSC cannot form). If the wavelength of the incident wave is much smaller than
the radius $R$ of the cylinders, that is, $\omega R\ll c_{t}
$, then 
variations of the field $u$ within each cylinder can be neglected 
so that the integral in (\ref{eq:LS}) is reduced
to the sum over cylinders:
\begin{equation}\label{eq:FF}
u_j(r)=u^{0}_j(r)+ \sum_{n}G_{ji}(r-n\hat{y})p_i(n)
\end{equation}
where $p(n)=\epsilon P(r_n)$ is the induced dipole moment of the 
cylinder at $r_n=(0,n)$,
$n=0,\pm 1,\pm 2,...$, and $\epsilon=\pi R^{2}$ is the area of the cross section
of the cylinder.

To simplify calculation of the induced dipoles (\ref{eq:2}),
 the lame coefficients of the cylinders and the background media 
are assumed to be the same, $\xi_{\lambda,\mu}=0$. In this case, 
$$
p_j(n)=- \omega^{2}\epsilon \xi_{\rho}u_j(n\hat{y})=- 
\omega^{2}\epsilon \xi_{\rho}u_j(0)e^{ik_{y}n}
$$
where the Block periodicity of $u_j(r)$ was used.
Setting $r=0$ in (\ref{eq:LS}), 
a consistency condition on the the induced dipole moment of the central scatter, $p(0)$,
is obtained,  
which, in turn, determines the field on the central scatter:
\begin{equation}
\label{eq:IP}
u_{i}(0)=\frac{u^{0}_{i}(0)}{1+\omega^{2}\xi_{\rho}\sum_{n}e^{ik_{y}n}\int_{\mid s \mid<R}d^{2}sG_{ii}(s+n\hat{y})}\,.
\end{equation}
no summation over repeated indices here. The off-diagonal components of 
$G_{ji}$ do not contribute to the right side of (\ref{eq:LS}) when we evaluate $u_{i}(0)$.
These components determine the
 contribution to
the $\hat{x}$ ($\hat{y}$) component of the scattered field 
on the central scatter coming from the $\hat{y}$ ($\hat{x}$) component of the
radiation field produced by the induced dipoles of all the other scatters.
The off diagonal terms, $G_{xy}(r)=G_{yx}(r)=\frac{i}{4\omega^{2}}\nabla_{x}\nabla_{y}[H_{0}^{(1)}(k_{l} |r|)-H_{0}^{(1)}(k_{t} |r|)]$, contain a second partial derivative, 
$\nabla_x\nabla_y$, acting on a radially symmetric function, the linear combination of 
$H_{0}^{(1)}(k_{a} |r|)$. Therefore, the contribution of these terms 
to the right side of (\ref{eq:LS}) 
is odd in $x$, and hence it vanishes along the line $x=0$ (along the array).
This is necessary for the continuity of the displacement field.

Using the asymptotic expansion of the Hankel function for large values of the 
argument, the series in (\ref{eq:FF}) can be written as a linear 
combination of plane waves in the asymptotic regions $|x|\to \infty$ (by 
neglecting the evanescent (exponentially decaying) modes). The propagating 
modes are proportional to $e^{ik_{a,x}|x|}$ where $k_{a,x}>0$. In the spectral range under consideration, there are only two open diffraction 
channels, one for each polarization $a$. The amplitude of the scattered wave in the asymptotic region 
$x\to-\infty$ is given by $u^R_{a}=R_{ab}(\omega)u^{0}_{b}$, where $R_{ab}$ is the reflection matrix, whereas $u^T_{a}=T_{ab}(\omega)u^{0}_{b}$ is the amplitude 
of the transmitted wave
when $x\to \infty$. Since all induced dipoles are proportional to $u_j(0)$,
the frequency dependence of the reflection (or transmission) matrix $R(\omega)$ (or $T(\omega)$)
is fully determined by the frequency dependence of $u_j(0)$. In the complex
plane $\omega^2$, $R(\omega)$ and $T(\omega)$ have a pole if the scattering structure has a 
resonance. Near the pole the reflection matrix has the following form:
\begin{equation}
\label{eq:R}
  R(\omega) \sim \frac{\tilde{R}}{\omega^{2}-\omega_{0}^{2}+i\Gamma}+K_{0}\,,
\end{equation}
 where $\tilde{R}$  is the residue matrix,
 and
$K_{0}$ is the analytical part of $R(\omega)$ evaluated at the pole (it describes the so called background
scattering), the transmission matrix 
$T(\omega)$ has a similar form near a resonance. 
Note that the graph of
$|u^R_{a}|$ has the standard Lorentzian shape as a function of real $\omega^2$ for both in-plane polarizations.

\begin{figure}[!htb]
\minipage{0.4\textwidth}
   \includegraphics[width=3in,height=1.5in]{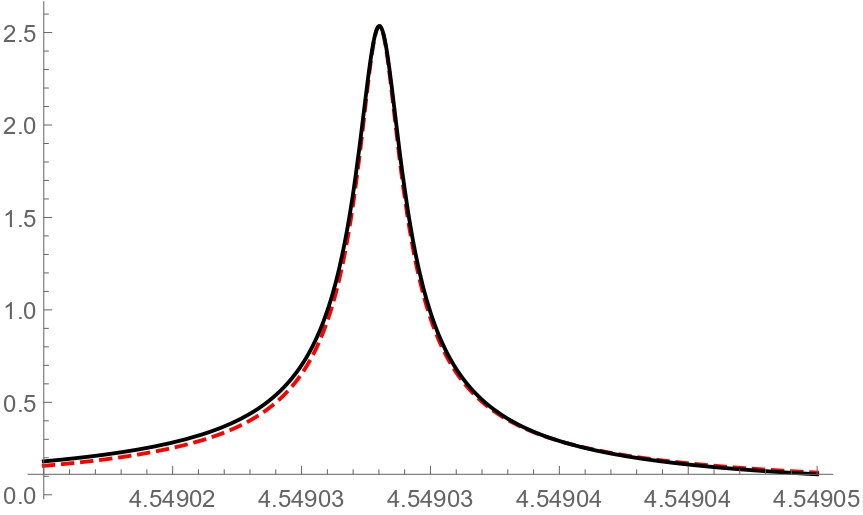}

\endminipage\hfill
\minipage{0.4\textwidth}
   \includegraphics[width=3in,height=1.5in]{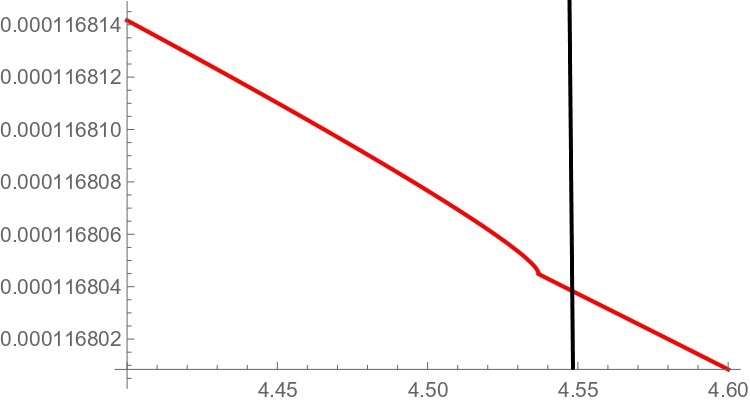}
  
\endminipage
\caption{\small Left panel: Plot of $\mid \frac{\epsilon u_{x}(0)}{u_{x}^{0}(0)}\mid$ vs $s=\omega/c_{t}$ for the following parameter values: $(k_{y},\epsilon,\xi_{\rho},\alpha)=(1.73405,.01,.28,.59915)$. The solid black line correspond to the exact solution and the red dashed line corresponds to the Lorentzian fit with fitting parameters $(\omega_{0},\Gamma)=(4.54904c_{t},6.62285 \cdot 10^{-6}c_{t})$;
Right panel: Plot of $\frac{\epsilon u_{y}(0)}{u_{y}^{0}(0)}$ vs $s$ for the same parameter values as the left panel, the black vertical line corresponds to the resonance position specified by the Lorentzian fit.}
\end{figure}

The solid black line in 
Figure 1 shows $\frac{\epsilon |u_{x}(0)|}{|u^{0}_{x}(0)|}$ (left panel) and $\frac{\epsilon |u_{y}(0)|}{|u^{0}_{y}(0)|}$ (right panel) calculated from 
(\ref{eq:IP}) as a function 
of $s=\omega/c_t$. The red dashed line shows a numerical fit to the theoretical
curve by a Lorentzian profile (the resonance frequency $\omega_0$, width
$\Gamma$, and the 
maximal value are the fitting parameters). Two important observations
follows from this numerical analysis.  First, the contribution 
of $u_y(0)$ to the induced dipole is negligible (right panel) as 
$|u_x(0)|/|u_y(0)|\sim 10^{-4}$ (if $|u^0_x|\sim |u_y^0|$), and
 it contributes only to the background 
scattering. Second, the scattering is dominated by a resonance (left panel)
so that the background scattering $K_0$ can be neglected
near the resonance in (\ref{eq:IP}).

\begin{figure}[!htb]
\minipage{0.2\textwidth}
   \includegraphics[width=3in,height=1.5in]{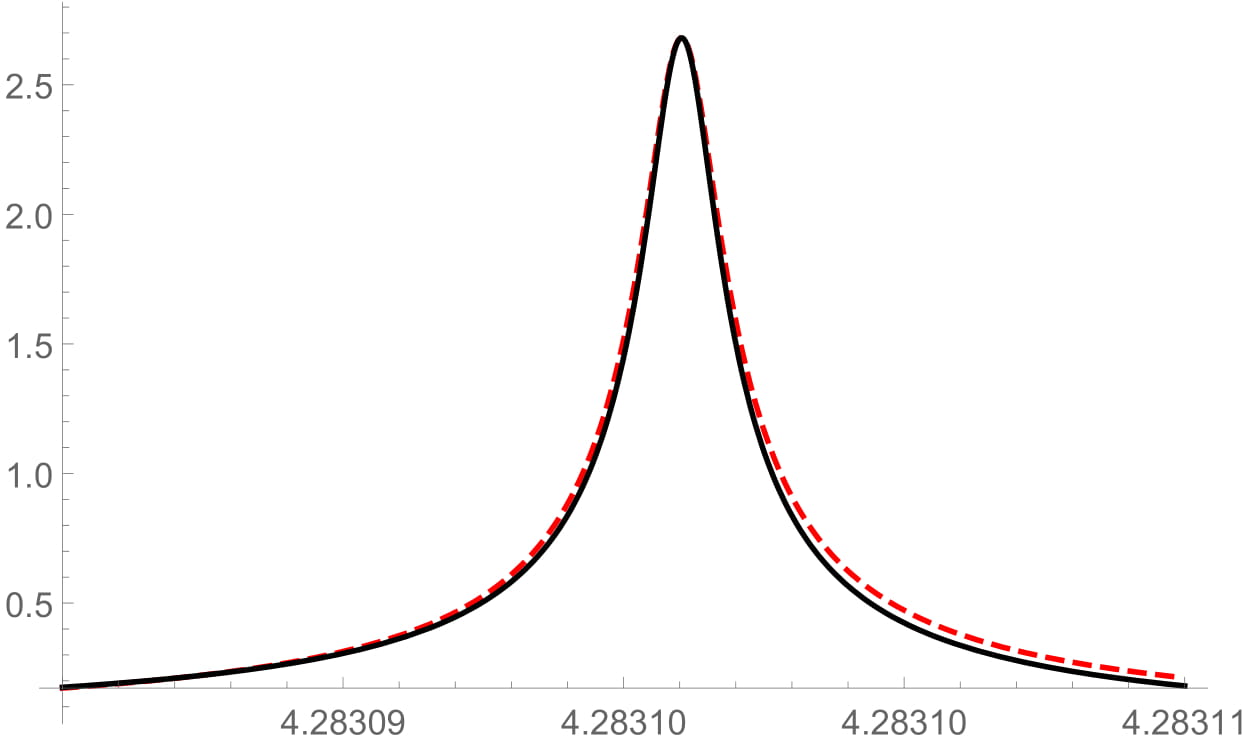}
\endminipage\hfill
\minipage{0.4\textwidth}
   \includegraphics[width=3in,height=1.5in]{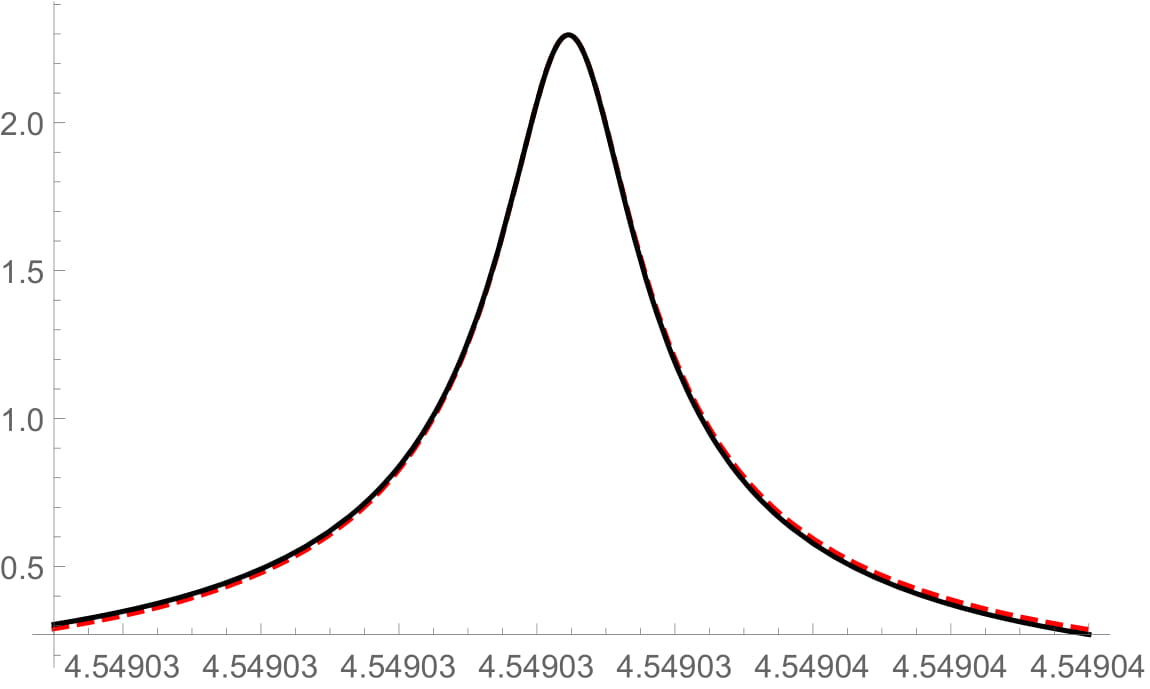}
\endminipage\hfill
\caption{\small Left panel: Plot $\mid \frac{\epsilon u_{x}(0)}{u_{x}^{0}(0)}\mid$ vs $s$ for $(k_{y},\epsilon,\xi_{\rho},\alpha)=(2,.01,.28,.59915)$ the fitted parameters of the Lorentzian profile are $(\omega_{0},\Gamma)=(4.2831c_{t},6.07586 \cdot 10^{-6}c_{t})$.; Right panel: Plot $\mid \frac{\epsilon u_{x}(0)}{u_{x}^{0}(0)}\mid$ vs $s$ for $(k_{y},\epsilon,\xi_{\rho},\alpha)=(1.73405,.01,.28,.65)$ the fitted parameters of the Lorentzian curve are $(\omega_{0},\Gamma)=(4.549032c_{t},8.63167 \cdot 10^{-6}c_{t})$.}
\end{figure}

\begin{figure}[!htb]
\minipage{0.2\textwidth}
   \includegraphics[width=3in,height=1.5in]{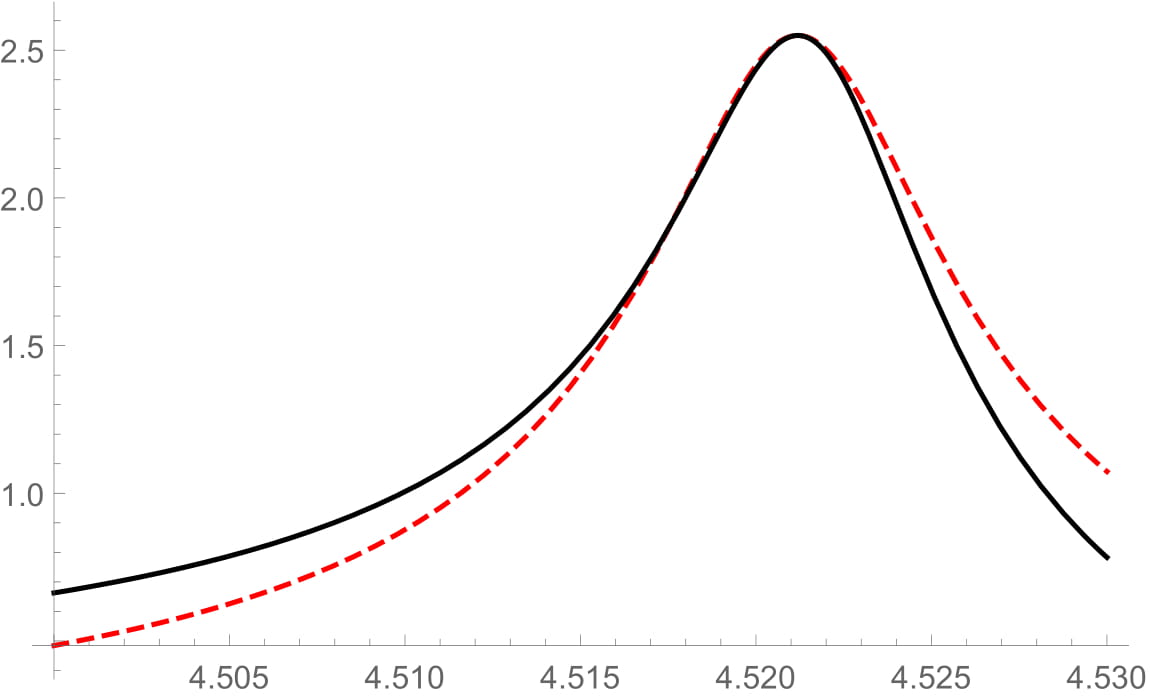}
\endminipage\hfill
\minipage{0.4\textwidth}
   \includegraphics[width=3in,height=1.5in]{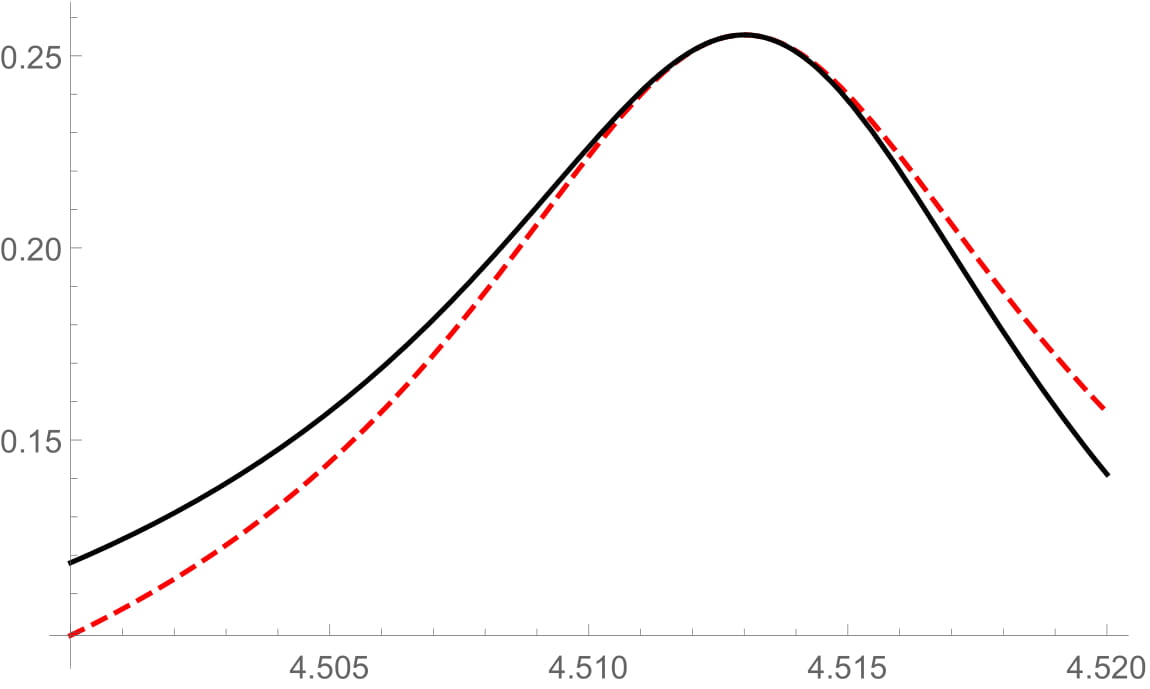}
\endminipage\hfill
\caption{\small Left panel: Plot $\mid \frac{\epsilon u_{x}(0)}{u_{x}^{0}(0)}\mid$ vs $s$ for $(k_{y},\epsilon,\xi_{\rho},\alpha)=(1.73405,.1,.28,.59915)$ the fitted parameters of the Lorentzian profile are $(\omega_{0},\Gamma)=(4.52119c_{t},.037c_{t})$.; Right panel: Plot $\mid \frac{\epsilon u_{x}(0)}{u_{x}^{0}(0)}\mid$ vs $s$ for $(k_{y},\epsilon,\xi_{\rho},\alpha)=(1.73405,.01,2.8,.59915)$ the fitted parameters of the Lorentzian curve are $(\omega_{0},\Gamma)=(4.513c_{t},.0493664c_{t})$.}
\end{figure}

The conclusion holds for significant variations of the system parameters,
although the parameters of the Lorentzian profile may change significantly.
Figures 2 and 3 display  $\frac{\epsilon |u_{x}(0)|}{|u^{0}_{x}(0)|}$ 
and its fit by a Lorentzian profile for various values 
of the parameters  $(k_{y},\epsilon,\xi_{\rho},\alpha)$ (indicated in the captions). One can see that the resonance remains relatively narrow under increasing $\alpha$ or $k_{y}$, while increasing 
$\epsilon$ and $\xi_{\rho}$ by
an order of magnitude results in a drastic change in the resonance width
(as might be seen in Fig.~4).
It should be noted, however, that as $\epsilon$ and $\xi_{\rho}$ increase, the 
dipole approximation becomes inapplicable so that Eq. (\ref{eq:IP}) is no longer valid and one must solve (\ref{eq:LS}) by other means. 
 As a point of fact, that the scattering in the system considered 
is resonance dominated
 can be proved analytically without all the simplifying 
assumptions made above. However mathematical details of this study 
are rather involved and are omitted. 

\subsubsection{\small A Fabry-Perot interferometer with two propagating coupled
modes} 

Consider a Fabry-Perot resonator made of two resonating scattering interfaces
such as the one discussed above. 
If the distance between the interfaces is large enough so that the evanescent fields
near each interface do not contribute to the field on the other interface, then
the reflection matrix of this composite structure is:
\begin{eqnarray}\nonumber
  R_{FP}(\omega)&=&R(\omega) +T(\omega)D(\omega,d)R(\omega)D(\omega,d)\\
\label{eq:RFP}
&&\times(I-(R(\omega)D(\omega,d))^{2})^{-1}T(\omega)\\ \nonumber
T_{FP}(\omega)&=& T(\omega)D(\omega,d)(I-(R(\omega)
D(\omega,d))^{2})^{-1}T(\omega)\,,
\end{eqnarray}
where $I$ is the unit diagonal matrix,
$D(\omega,d)$ is the matrix defined by the amplitude of the wave 
after propagation through a distance $d$, in this case $d$ is the distance 
between the interfaces. It is a diagonal matrix with elements being 
phase factors corresponding to different group velocities of the modes. 
A resonance position of the composite system (a pole of $R_{FP}(\omega)$)  
is defined by the condition
\begin{equation}\label{eq:res0}
 \det[I-(R(\omega)D(\omega,d))^{2}]=0\,.
\end{equation}
The scattering on each interface is further assumed to be  resonance dominated (the background scattering can be neglected). This assumption
is always justified if
the scattering interface is composed of small identical (subwavelength) scatters
separated by distances that are much larger than the size of scatterers
as in the example presented above.
Then the analytic part of the single interface reflection matrix 
will scale with the volume of the scatter so that the background scattering 
$K_0$
can be neglected in (\ref{eq:R}) as compared to 
the resonant part. For the array considered
above this approximation holds because $u_{y}(0)$ is negligible as compared 
to $u_x(0)$.
In this case, Eq. (\ref{eq:res0}) is further simplified to
\begin{equation}\label{eq:res}
 \det[(\omega^{2}-\omega_{0}^{2}+i\Gamma)^{2}I-(\tilde{R}D(\omega_{0},d))^{2}]=0
\end{equation}
If the polarization modes are decoupled ($\tilde{R}$ is diagonal), then 
this equation is reduced to  
 the case of light scattering on a dielectric double array  \cite{PRL2008}
where it was shown that, for specific values 
of the distance $d$, there exists a real root $\omega^2$ that lies in the radiation continuum, thus 
indicating a BSC as a resonance with the vanishing width. Here 
the situation is more complicated as the residue matrix $\tilde{R}$
is not diagonal due to the coupling of the compression and sheer modes,
and the phase factors in $D$ are different for each mode because $c_t\neq c_l$.
 
\subsubsection{BSC in a double array of cylindrical scatters}
\setcounter{equation}0

The existence of BSC depends on the structure of the residue 
matrix $\tilde{R}$. Its calculation generally requires solving the Lippmann-Schwinger integral equation. If the dipole approximation
is applicable, this task is simplified to summation of the dipole radiation.
The latter can be done analytically for either a single or double
periodic array of thin, long, cylindrical scatter where the 
longitudinal and transverse polarizations in the plane perpendicular
to the cylinders are coupled at the surface of the cylinders. 
As noted, the transverse mode polarized parallel to the cylinders is decoupled.
Therefore the three-dimensional reflection matrix is block-diagonal.
The  $1\times 1$ block corresponds to the decoupled transverse mode.
A study of BSCs of this mode 
is essentially identical to the electromagnetic BSCs in one 
or more open diffraction channels \cite{PRL2008,JMP2010}.

The $2\times2$ block describes two coupled polarization modes
(the in-plane longitudinal and transverse modes)
in the plane perpendicular to the cylindrical scatters. 
The poles of the scattering matrix are defined by roots of (\ref{eq:res}).
The residue matrix is calculated using the asymptotic expansion 
of the Hankel function in (\ref{eq:LS}) and the Poisson summation formula
in (\ref{eq:res}) (much like in the electromagnetic case \cite{JMP2010}). Recall that only a special case is considered  
in which 
the cylinders have different density compared to the background material but the lame coefficients remain the same. Materials with such properties can be found
in \cite{materials}. 
 In order to obtain the reflection coefficients one can change coordinates to the in-plane unit polarization vectors:
$\hat{e}_{l}=\hat{k}_{l}=(k_{l,x}\hat{x}+k_{y}\hat{y})/k_l$ and
$\hat{e}_{t}=-(\hat{k}_{t} \times \hat{z})=-(k_{y}\hat{x}-k_{t,x}\hat{y})/k_t$,
where $k_{a,x}=\sqrt{k_{a}^{2}-k_{y}^{2}}$. 
The position of the resonance pole ($\omega_0^2$ and 
$\Gamma$) is calculated in the leading 
order of $\epsilon$.
Expanding $u_{x}(0)$ to leading order in $\epsilon$ one infers  that
\begin{eqnarray*}
 &&\tilde{R}_{ll}=p_{l,x}\frac{i\Gamma}{p_{l,x}+k_{y}^{2}p_{t,x}^{-1}}\,,\quad
 \tilde{R}_{lt}=-\alpha k_y\frac{i\Gamma}{p_{l,x}+k_{y}^{2}p_{t,x}^{-1}}\,,\\
 &&\tilde{R}_{tt}=\frac{k_y^2}{p_{t,x}}\,\frac{i\Gamma}{p_{l,x}+k_{y}^{2}p_{t,x}^{-1}}\,,\quad
 \tilde{R}_{tl}=-\frac{p_{l,x}k_y}{\alpha p_{t,x}}\,\frac{i\Gamma}{p_{l,x}+k_{y}^{2}p_{t,x}^{-1}}
\end{eqnarray*}
where $\Gamma=\epsilon\xi_\rho \beta\omega_1^2 (p_{l,x}+k_{y}^2p_{t,x}^{-1})$, $\beta=\frac 14\epsilon^2\xi_\rho^2(k_y-2\pi)^2$,
$p_{l,x}=\sqrt{\alpha^{2}(k_{y}-2\pi)^{2}-k_{y}^{2}}$, 
and $p_{t,x}=\sqrt{(k_{y}-2\pi)^{2}-k_{y}^{2}}$.

It should be noted that $\tilde{R}_{lt},\tilde{R}_{tl} \neq 0$, 
as a result the two in-plane polarization modes are coupled
at the interface. It follows from the structure of $\tilde{R}$ that
$\det(\tilde{R})=0$. Physically, this is because near the resonance frequency, the elastic field on the scatter lies almost completely parallel to the $\hat{x}$ direction, this means that the reflection amplitudes, $u_{l}^{R},u_{t}^{R}$, are  proportional to $u_{x}(0)$ because $u_{y}(0)$ is of higher order in the volume of the scatters (as shown by the right panel of Fig. 1):
$$
u_{l}^{R},u_{t}^{R} \propto u_{x}(0) \propto u_{x}^{0}(0) \propto u_{l}^{0}\hat{e}_{l,x}+ u_{t}^{0}\hat{e}_{t,x}
$$
This feature of the scattering process is easily understood in the considered
dipole approximation. In this case the induced dipole moment 
of each scatterer is parallel to $\hat{x}$ by the reflection symmetry
about the position of each scatterer
so that the rows of the reflection matrix are linearly dependent.
Therefore, neglecting $K_{0}$, 
$\det(\tilde{R})= \det(R(\omega))=0$ near the resonance frequency. Equation (\ref{eq:res}) can be viewed as an eigenvalue problem for the 
$2\times2$ matrix $\tilde{R}D(\omega_{0},d)$. Since $\det (\tilde{R}D(\omega_{0},d))=
\det(D(\omega_{0},d))\det(\tilde{R})=0$, $\omega^2=\omega_0^2-i\Gamma$ is always a solution. However, it does not define a pole in (\ref{eq:RFP}) if 
$R(\omega)$ has the form (\ref{eq:R}) (all singular factors 
$(\omega^2-\omega_0^2+i\Gamma)^{-1}$ are cancelled out, this can be seen explicitly by using the unitary condition on the scattering matrices). The other roots 
determine the poles $\omega_\pm^2$ of the composite structure:
\begin{eqnarray}
 \omega_{\pm}^{2}-\omega_{0}^{2}+i\Gamma&=&\pm \sqrt{{\rm Tr}(\tilde{R}D(\omega_{0},d))^2}\nonumber\\ \label{eq:newpoles}
&=& \pm i\Gamma\,
\frac{p_{l,x}e^{ip_{l,x}d}+k_{y}^{2}p_{t,x}^{-1}e^{ip_{t,x}d}}
{p_{l,x}+k_{y}^{2}p_{t,x}^{-1}}\,,
\end{eqnarray}
 A BSC occurs if the imaginary part of the pole
can be driven to zero by adjusting parameters of the system. 
This happens 
 if the phase factors in the numerator in 
the right side of  (\ref{eq:newpoles}) become unit, $e^{ip_{l,x}d}=
e^{ip_{t,x}d}=\pm 1$, that is,
both modes satisfy the quantization conditions:
$$
 p_{l,x}d=\pi M<p_{t,x}d=\pi N
 $$
where $M$ and $N$ are both either mutually even or odd integers.
It should be noted that the system has a parity symmetry in the $x$ direction. The even/odd integers in the phase factors above corresponds
to even/odd parity BSC just as in the scalar case. In contrast to electromagnetic case \cite{PRL2008}, the quantization 
conditions cannot be satisfied by adjusting the distance $d$ between 
the arrays because the modes have different dispersions 
(note the parameter $\alpha$ in $p_{l,x}$).  
 Both conditions can only be fulfilled for a generic 
parameter $\alpha$ if the spectral parameter $k_y$ is such that 
$$
 0<\frac{p_{l,x}}{p_{t,x}}=\frac{n}{m}<1
$$ 
for some integers $n$ and $m$. It follows from (\ref{eq:newpoles}) that
under the stated conditions one of the resonances turns into a BSC
at the resonance position of the single array
$$
 \omega_{BSC}^{2}= \omega_0^2
$$
while the other resonance has a double width $2\Gamma$ and the same position
$ \omega_{BSC}^{2}$,
just like in the electromagnetic case.

A BSC is not coupled to radiation modes, hence, cannot be excited 
by incident waves. 
It is a solution to the homogeneous equation (\ref{eq:LS}) for
the double array that is square integrable in $x$ and Bloch-periodic
in $y$ (Sommerfeld conditions at infinity $|x|\to \infty$ are satisfied
due to square integrability). 
If the cylinders in a double 
array are located at $r=r^\pm_{n}=\pm \frac 12d\hat{x}+n\hat{y}$, $n$ 
being an integer, then 
a BSC has a real frequency that lies in the radiation continuum, $\omega^2=\omega^2_{BSC}>c_{a}^{2}k_{y}^{2}$.  Just as in the case of the single array, the scattering matrices are determined by the dipole moment of the central scatter of each array $\tilde{u}_i(r^\pm_0)$, ($\tilde{u}(r)$ is the elastic field of the double array). The consistency condition on the dipole moment of each central scatter (located at $r^{\pm}_{0}$) results in a system of equations which has 
a non-trivial solution only if its determinant vanishes, which defines
$\omega_{BSC}$.

It was shown that in the Fabri-Perot limit, 
$k_{a,x}d \gg 1$ for every polarization state $a$,
such frequency does exist and
$\omega^2_{BSC}=\omega_0^2$, where
$\omega_0$ is the resonance position for a single array if the distance $d$ between 
arrays and $k_y$ satisfy the conditions stated above.
The BSC field $\tilde{u}(r)$ is given by a similar expression as the scattered field in the right side 
of (\ref{eq:FF}), with necessary modifications in order to account for the second array. By construction of the induced dipoles $\tilde{u}(r_0^\pm)$
(without an incident wave), 
the amplitudes
of propagating modes in (\ref{eq:FF}) should vanish in the far field when 
$\omega^2=\omega^2_{BSC}$ (by the energy flux conservation) so that
a BSC solution $\tilde{u}(r)$ to the double array decays exponentially 
\begin{equation}\label{eq:BSC}
 \mid \tilde{u}(r) \mid \sim e^{-\frac{\epsilon \mid \xi_{\rho} \mid (k_{y}-2\pi)^{2} \mid x \mid}{2}}
\end{equation}
as $|x|\to\infty$,
and, hence, has a finite energy per
each period of the array. 

The energy density of a BSC 
is shown 
in the middle panel of Figure 4 in unit of $\frac{\rho_{b} c_{t}^{2}\mid \tilde{u}_{x}(\frac{d}{2}\hat{x}) \mid^{2}}{2}$. 
The left panel shows ${\rm Re}[\frac{\tilde{u}_{x}(r)}{\tilde{u}_{x}(\frac{d}{2}\hat{x})}]$.
It is also worth noting that the $y-$component 
of the BCS solution,
$\tilde{u}_{y}(r)\sim O(\epsilon)$, so its contribution to the energy density is of higher order in $\epsilon$ and can be neglected. The parts of the array shown in the middle 
and right panels corresponds to a region indicated by a rectangle in 
the left panel where a pictorial representation of the array is shown.
As one can see, the energy and the field is concentrated near 
each array.
The displayed BSC has an even parity
so that $\tilde{u}_{x}(r)$ is odd in $x$ and  
$ d\sqrt{k_{BSC}^{2}-(k_{y}-2\pi)^{2}} \gg 1$,
so that the discussed 
Fabry-Perot limit is satisfied. The material parameters correspond to bulk silicon. According to \cite{materials}-\cite{materials2}, the relative density between different samples can be tuned to at least an order of magnitude larger than the relative lame coefficients to justify the simplifying assumption about the lame 
coefficients. So, the found BSC can also be observed experimentally. 
\begin{figure}[!htb]
\minipage{0.2\textwidth}
   \includegraphics[width=1in,height=1in]{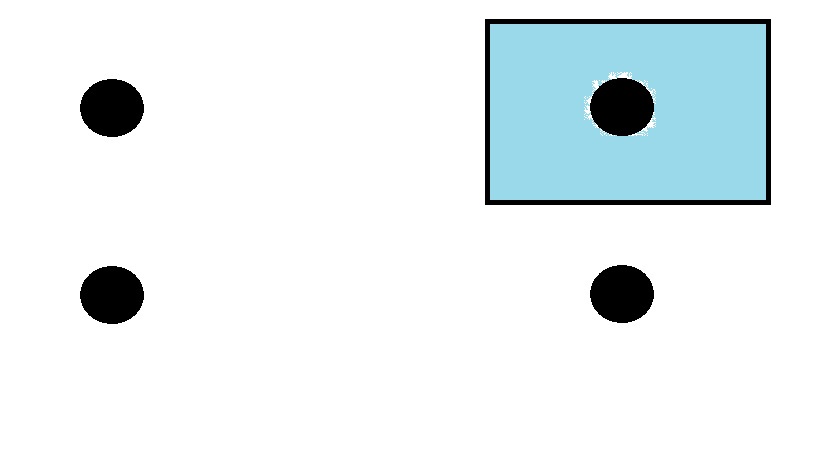}
\endminipage\hfill
\minipage{0.4\textwidth}
   \includegraphics[width=3in,height=1.5in]{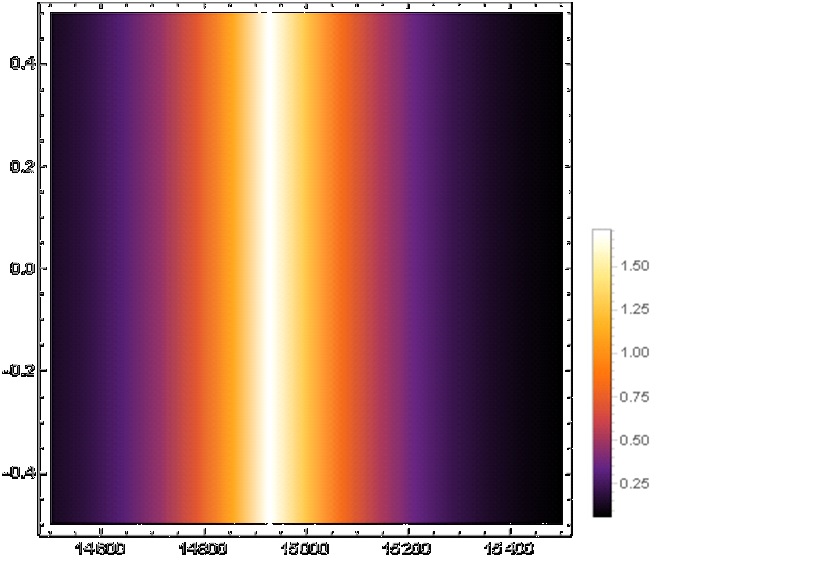}
\endminipage\hfill
\minipage{0.4\textwidth}%
   \includegraphics[width=3in,height=1.5in]{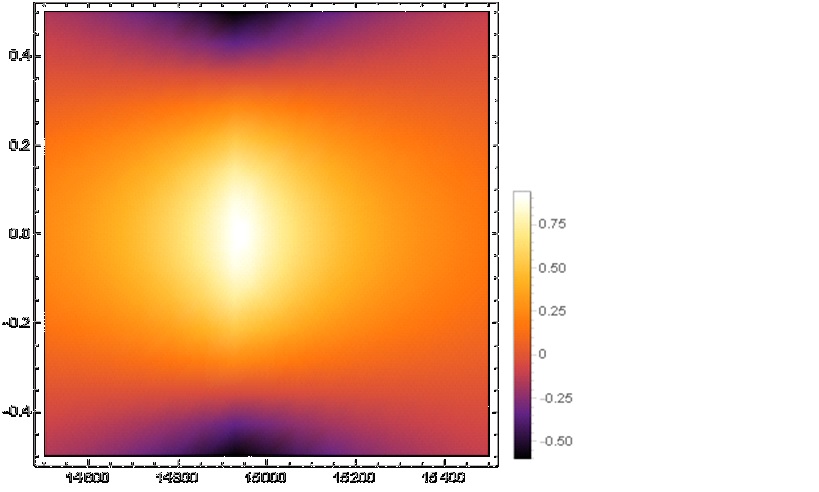}
\endminipage
\caption{\small Left panel: Schematic of the double array of periodically positioned elastic
cylinders, the $x-$axis is horizontal, the $y-$axis is vertical; Middle panel: The energy density of a BSC in a vicinity of the right arrays for parameters $(k_{y},\epsilon, d,\xi_{\rho},\alpha)=(1.73405,.001,29880,.280,.59915)$, the frequency of this BSC is given by $\omega_{BSC}=4.549134c_{t}$. The vertical grid is defined in the text and the horizontal scale is measured in units of the period of the array;
Right panel: The $x-$component of the displacement field of this BSC, the grid legend and parameters are identical to the former.}
\end{figure}
 
In the frequency range under consideration, the two polarizations decouple for normal incident $k_y=0$, the scattering matrix is proved to
 become diagonal, that is, 
the longitudinal and transverse wave are decoupled in the first open diffraction
channel (which can be understood from the parity symmetry in $x$ direction).
 The transmission coefficient for the transverse mode
becomes $T_{tt}(\omega)=1-O(\epsilon)$, which can be realized from the induced dipole radiation analysis presented above.
It is therefore clear that a transverse BSC cannot exist
in this spectral range. However the longitudinal mode
has a resonance, and
the longitudinal reflection coefficient 
 is shown to have the Breit-Wigner form near this resonance, hence, 
BSCs exist. These are single-mode BSCs and have already been discussed in many contexts.
For higher diffraction channels the longitudinal and transverse modes 
are coupled even for $k_y=0$. However, the analysis of BSCs in higher 
diffraction channels 
 becomes mathematically involved even in the single mode case 
\cite{JMP2010} and will not be given here.

\subsubsection{Conclusions}

It was shown that elastic meta-interfaces can be used to obtain
BSC via tuning of the cavity width and Bloch phase. Elastic BSC are shown to contain coupled transverse and longitudinal 
modes that have different dispersion relations. In contrast to 
previously reported BSC, such as the case of the electromagnetic double array, this elastic BSC requires very peculiar condition on the phases of both the transverse and longitudinal phase, such conditions are necessary in order to tune the far field radiation to zero. This fine tuning is an interesting artifact of the elastic double array which is a major distinction from the BSC previously studied in photonic and acoustic systems. 
An analytic solution is obtained for such 
BSC for the first time. Although the stated results have been proved under simplifying assumptions ($\xi_{\lambda,\mu}=0$ and the Fabri-Perot limit), 
the same dipole formalism can be used in the general case, however this is mathematically involved and will be discussed in a subsequent publication. 

Elastic BSCs can be used as elastic wave guides or as resonators 
with high quality factors  in a broad spectral
range, especially in view of that elastic systems supporting BSC 
can be designed using mechanical metamaterials (as materials with 
desired elastic properties). In particular, owing to a high sensitivity 
of the quality factor to geometrical and physical properties of 
a resonating system, elastic BSCs can be used to detect 
impurities in solids from variations of the density. Understanding the parallels and distinction among wave phenomena in a variety of physical systems is necessary for construction of sensor, filters, lasers, etc. This sort of artificial wave construction will allow one to analyze non-linear phenomena in elastic material in a way similar to 
the studies done in photonics \cite{Remy2013}.  The energy density 
of a high quality resonance (near-BSC state) has shown to exhibit ``hot'' spots where
it exceeds the energy density of the incident wave by orders in magnitudes and would allow one to amplify non-linear effects in solids in a controlled manner.

\end{document}